\documentclass{sig-alternate-no-copy}
\usepackage[latin1]{inputenc}
\usepackage[english]{babel}
\usepackage{balance}
\usepackage{times}

\usepackage{color}

\definecolor{darkgreen}{rgb}{0,.5,.2}
\definecolor{gray}{rgb}{.5,.5,.5}

{\vspace{5mm}\begin{sloppypar}
  \noindent\hrulefill\, BEGIN \,\hrulefill
  \end{sloppypar}
}%
{\begin{sloppypar}
  \noindent\hrulefill\, END   \,\hrulefill
  \end{sloppypar}\vspace{5mm}
}

\newcounter{todo}

\newcounter{question}

\newcounter{comment}

\begin{document}

\title{Towards a Query Language for the Web of Data \\(A Vision Paper)}


\numberofauthors{2}
\author{
 \alignauthor Juan Sequeda\\
       \affaddr{University of Texas at Austin}\\
       \email{jsequeda@cs.utexas.edu}
 \alignauthor Olaf Hartig\\
       \affaddr{Humboldt-Universit\"at zu Berlin}\\
       \email{hartig@informatik.hu-berlin.de}
}

\maketitle              

\renewcommand{\thefootnote}{\fnsymbol{footnote}}
\setcounter{footnote}{0}

\begin{list}{}%
  {\setlength{\leftmargin}{19mm}}%
         \item[]%
	\textit{``You want a good thesis? IR is based on precision and recall and the minute you add semantics, it is a meaningless feature. Logic is based on soundness and completeness. We don't want soundness and completeness. We want a few good answers quickly.''} -- Prof.~James A.~Hendler, 2009, on the topic of answering queries over the Semantic Web.\footnote{http://www.youtube.com/watch?v=sbmMxzOeZ-4}
\end{list}

\renewcommand{\thefootnote}{\arabic{footnote}}
\setcounter{footnote}{0}

\section{Introduction} \label{Section:Intro}
As of today, 2011, the Web of Data is composed of RDF based datasets that are exposed on the World Wide Web i) in adherence to the Linked Data principles, ii) using the SPARQL protocol \emph{or} iii) as static RDF documents. The Web of Data is in constant growth. We foresee that it will become more wild, uncontrollable and infinite. It will have no boundaries and will grow faster than it can be crawled. It will be gigantic ... and we want to query it!

What do we mean by ``querying the Web of Data''? From the current Web search paradigm, it could mean that we first crawl the data, index it, and then search based on keywords over the indexed data. From a data warehouse perspective, it could mean to copy relevant datasets into a local RDF database and execute queries over the local collection. Another approach would be to execute declarative queries on the fly over the Web of Data itself. In this vision paper, we focus on the latter because it entails new challenges and open questions, which we will describe in this paper. 

The main question is, what should a declarative language for such an approach be?
Since the Web of Data is based on the RDF model and SPARQL is the standard query language for RDF, it seems natural to ask: Is SPARQL suitable as a declarative language to query the Web of Data? 
The semantics of SPARQL, as given in the current standard, considers a single, fixed, a priori defined RDF dataset. It has been defined in the context of databases and logic and query results are assessed based on the concept of soundness and completeness. However, to query the Web of Data, we would need query language that considers an unbounded, distributed collection of RDF data which cannot be assumed to be known completely.
What characteristics of the Web of Data should be considered in order to define such a query language?

In the remainder of this paper, we first present related work and argue that research on querying the Web of Data is still in its infancy.  We then provide an initial set of general features that we envision should be considered in order to define a query language for the Web of Data. Furthermore, for each of these features, we pose questions that have not been addressed before in the context of querying the Web of Data. We believe that addressing these questions and studying these features may guide the next 10 years of research on the Web of Data.
\section{Related work} \label{Section:RelWork}
Research on querying the World Wide Web started in the mid 1990s~\cite{Florescu98:DBTechniquesForWWW}. It is important to note that the Web at that time only consisted of linked hypertext documents. Most of the research was based on developing models to represent the Web (e.g.~\cite{Mendelzon98:FormalModelsOfWebQueries,Abiteboul00:QueriesAndComputationOnTheWebArticle,Kleinberg99:WebAsGraph}) and approaches for (declarative) queries over the Web (e.g.~\cite{Konopnicki95:WWWQuerySystemW3QS,Mendelzon97:QueryingTheWWW,Spertus00:StructuredQueryLanguageForTheWeb}). To the best of our knowledge, the last paper published on this topic was in 2002~\cite{Spielmann02:DistributedWebQuerying}. Four years later, Tim Berners-Lee proposed the Linked Data principles, which kick-started the Linked Open Data project. This project helped to bootstrap the Web of Data as it exists today. The first paper that explicitly focused on querying the Web of Data was published in 2009~\cite{Hartig09:QueryingTheWebOfLD}. Since then,
	further papers
have been published (e.g.~\cite{Bouquet09:QueryingWebOfData,Harth10:DataSummariesForLDQueryProcessing,Ladwig10:LinkedDataQueryProcessingStrategies,Hartig10:DBPerspectiveOnConsumingLD,Schwarte11:FedX,Acosta11:ANAPSID}).
We believe that these works are only the beginning of a new area of research for which we aim to provide inspiration with this paper.

Other fields that should be considered relevant in this context, are distributed databases, uncertain and probabilistic databases, data stream management, and Deep Web.
\section{Features} \label{Section:Features}

\noindent {\bf  Scope:} According to the current SPARQL standard, the scope of a query is a predefined RDF dataset. A query language for the Web of Data, should not have such a fixed scope; instead, it should take advantage of the openness and the unbounded nature of the Web. A basis for defining the scope of queries in this context is a model of the Web of Data. We ask ourselves:
\begin{itemize} \addtolength{\itemsep}{-0.5\baselineskip}
 \item What characteristics of the Web of Data are relevant for a data model that can be used as the foundation for a query language?
 \item How would such a model deal with the dynamic nature of the Web?
 \item Should such a model capture different approaches of exposing datasets on the Web?
 \item How can the scope of queries be restricted to a particular, declaratively defined portion of the Web of Data?
\end{itemize}

\noindent {\bf Language Expressiveness:} The expressiveness of a query language is characterized by the type of questions that can be asked
using the language. However, adding expressive power usually increases the computational complexity of a query language. This issue becomes even more important in the context of computing queries at Web scale. Hence, developing a query language for the Web of Data comprises the challenge of finding a trade-off between expressiveness and complexity. Since the answer to this problem may be different, depending on the usage scenario, we foresee the emergence of multiple approaches. We ask ourselves:
\begin{itemize}\addtolength{\itemsep}{-0.5\baselineskip}
 \item Should the language be concerned with record linkage and semantically overlapping vocabularies? Should the language deal with entity and vocabulary mappings?
 \item What operators should the language support? Which are unsuitable (e.g.~negation)?
 \item Could unsuitable operators be included by enabling users to declaratively bound the scope for them? How can such a bounded scope be declared in the queries (e.g.~based on namespaces, based on specific SPARQL endpoints, etc)?
 \item Should the query language consider the topology of the Web and allow users to specify path expressions for explicitly guiding link traversal based data discovery?
 \item Can provenance requirements be expressed in the
language?
 \item Should the language be concerned with trustworthiness of data (or other criteria of information quality)? Could we make quality requirements explicit in queries?
\end{itemize}

\balance
\noindent {\bf Query Results:}
Queries are executed over collections of data in order to compute results that answer the questions expressed by the queries. In the context of the Web of Data it is not obvious what such an answer should be, because the data collection is unbounded and uncontrolled. Furthermore, some data (and thus query results) may not be considered trustworthy by certain users.
	On the other hand, personalized query semantics may emerge and query results could be influenced by the query history and behavior of a user's friends.
Thus, depending on the use cases we expect different types of results for the same query. Hence, we foresee multiple approaches for defining what a query result is. We ask ourselves:
\begin{itemize}\addtolength{\itemsep}{-0.5\baselineskip}
 \item Should query results be assessed based on soundness and completeness, precision and recall, a combination thereof, or even something else?
 \item If a query is executed multiple times, should the results be incremental?
 \item Should query semantics be monotonic or non-monotonic?
 \item Should query results depend on social aspects?
 \item Can parts of the Web of Data conceptually be locked during query execution? If not, what should the result be if the execution of a query uses data that might have already been altered by the time the execution terminates?
 \item Should query results include their provenance?
 \item Should query results be associated with trustworthiness scores?
\end{itemize}

\noindent {\bf Implementation Aspects:} Declarative queries can be computed in multiple ways, applying different execution strategies. Different query execution plans may be formed by combining alternative
	data
access paths and join algorithms. Query optimizers estimate costs for such plans in order to determine the most suitable plan. Such costs usually depend on I/O and statistics about the queried data. In the context of the Web of Data, such information may not be available or even relevant. On the other hand, new criteria become relevant (e.g.~network latency). Additionally, query semantics may require the discovery and exploitation of vocabulary mappings or entity mappings. Depending on the expressiveness of the query language, determining suitable query plans may become much more complex than it is in traditional query optimization scenarios. We foresee a significant focus on adaptive query processing instead of traditional optimize-then-execute, due to the lack of control on the Web of Data. We ask ourselves:
\begin{itemize}\addtolength{\itemsep}{-0.5\baselineskip}
 \item What is a logical and a physical execution plan for querying the Web of Data?
 \item Can optimization strategies developed in the database community be applied?
 \item How can a cost model be defined? What should it depend on?
 \item What type of statistics could be used to optimize queries?
 \item Can a query be optimized based on query plans from my friends' query engines?
 \item How can discovered data and intermediate results be indexed or cached?
 \item How can vocabulary mappings be found and used efficiently? 
 \item How can entity mappings (owl:sameAs links) be found and used efficiently?
\end{itemize}
\section{Closing Remarks} \label{Section:Conclusion}
We believe there is not a unique or even a right or wrong way of defining the features and answering the questions raised in this paper.
Instead, we envision researchers investigating and implementing different query languages for the Web of Data. If this is the case, another question arises: How do we evaluate and meaningfully compare different approaches?

To summarize, what should a query language for the Web of Data be? We do not know yet! However, we hope to have an answer to this question in the next 10 years.

\bibliographystyle{splncs}
\bibliography{main}

\begin{thebibliography}{10}

\bibitem{Florescu98:DBTechniquesForWWW}
Florescu, D., Levy, A.Y., Mendelzon, A.O.:
\newblock {Database Techniques for the World-Wide Web: A Survey}.
\newblock SIGMOD Record \textbf{27}(3) (1998)

\bibitem{Mendelzon98:FormalModelsOfWebQueries}
Mendelzon, A.O., Milo, T.:
\newblock {Formal Models of {W}eb Queries}.
\newblock Information Systems \textbf{23}(8) (1998)

\bibitem{Abiteboul00:QueriesAndComputationOnTheWebArticle}
Abiteboul, S., Vianu, V.:
\newblock {Queries and Computation on the Web}.
\newblock Theoretical Computer Science \textbf{239}(2) (2000)

\bibitem{Kleinberg99:WebAsGraph}
Kleinberg, J.M., Kumar, R., Raghavan, P., Rajagopalan, S., Tomkins, A.:
\newblock {The Web as a Graph: Measurements, Models, and Methods}.
\newblock In: Proc. of COCOON. (1999)

\bibitem{Konopnicki95:WWWQuerySystemW3QS}
Konopnicki, D., Shmueli, O.:
\newblock {W3QS: A Query System for the World-Wide Web}.
\newblock In: Proc. of VLDB. (1995)

\bibitem{Mendelzon97:QueryingTheWWW}
Mendelzon, A.O., Mihaila, G.A., Milo, T.:
\newblock {Querying the World Wide Web}.
\newblock Int. Journal on Digital Libraries \textbf{1}(1) (1997)

\bibitem{Spertus00:StructuredQueryLanguageForTheWeb}
Spertus, E., Stein, L.A.:
\newblock {Squeal: A Structured Query Language for the Web}.
\newblock Computer Networks \textbf{33}(1-6) (2000)

\bibitem{Spielmann02:DistributedWebQuerying}
Spielmann, M., Tyszkiewicz, J., den Bussche, J.V.:
\newblock {Distributed Computation of Web Queries Using Automata}.
\newblock In: Proc. of PODS. (2002)

\bibitem{Hartig09:QueryingTheWebOfLD}
Hartig, O., Bizer, C., Freytag, J.C.:
\newblock {Executing {SPARQL} Queries over the Web of Linked Data}.
\newblock In: Proc. of ISWC. (2009)

\bibitem{Bouquet09:QueryingWebOfData}
Bouquet, P., Ghidini, C., Serafini, L.:
\newblock {Querying The Web Of Data: A Formal Approach}.
\newblock In: Proc. of ASWC. (2009)

\bibitem{Harth10:DataSummariesForLDQueryProcessing}
Harth, A., Hose, K., Karnstedt, M., Polleres, A., Sattler, K.U., Umbrich, J.:
\newblock {Data Summaries for On-Demand Queries over {L}inked {D}ata}.
\newblock In: Proc. of WWW. (2010)

\bibitem{Ladwig10:LinkedDataQueryProcessingStrategies}
Ladwig, G., Tran, D.T.:
\newblock {Linked Data Query Processing Strategies}.
\newblock In: Proc. of ISWC. (2010)

\bibitem{Hartig10:DBPerspectiveOnConsumingLD}
Hartig, O., Langegger, A.:
\newblock {A Database Perspective on Consuming {L}inked {D}ata on the {W}eb}.
\newblock Datenbank-Spektrum \textbf{10}(2) (2010)

\bibitem{Schwarte11:FedX}
Schwarte, A., Haase, P., Hose, K., Schenkel, R., Schmidt, M.:
\newblock {FedX: Optimization Techniques for Federated Query Processing on
  Linked Data}.
\newblock In: Proc. of ISWC. (2011)

\bibitem{Acosta11:ANAPSID}
Acosta, M., Vidal, M.E., Lampo, T., Castillo, J., Ruckhaus, E.:
\newblock {ANAPSID: An Adaptive Query Processing Engine for SPARQL Endpoints}.
\newblock In: Proc. of ISWC. (2011)

\end{thebibliography}
\end{document}